\documentclass[reprint,showpacs,preprintnumbers,amsmath,amssymb,pre]{revtex4-1}
\usepackage{graphicx,wasysym}
\usepackage{dcolumn}
\usepackage{bm}
\usepackage{bbm}
\usepackage{natbib}
\usepackage{tabulary}
\usepackage{color}

\newcommand{\tobs}{t_N}

\newcommand{\Nw}{N_\mathrm{w}}
\newcommand{\Nc}{N_\mathrm{c}}

\newcommand{\W}{\mathcal{W}}
\newcommand{\Wun}{\mathbb{W}}
\newcommand{\app}[1]{\tilde{#1}}
\newcommand{\eigv}{\Xi}

\usepackage{mathrsfs}

\begin{document}

\title{Exact fluctuations   of nonequilibrium steady states \\from approximate auxiliary dynamics}
\author{Ushnish Ray}
\email{uray@caltech.edu}
\affiliation{%
Division of Chemistry and Chemical Engineering, California Institute of Technology, Pasadena, CA 91125
}%
\author{Garnet Kin-Lic Chan}
\email{garnetc@caltech.edu}
\affiliation{%
Division of Chemistry and Chemical Engineering, California Institute of Technology, Pasadena, CA 91125
}%
\author{David T. Limmer}
\email{dlimmer@berkeley.edu}
\affiliation{%
Department of Chemistry, University of California, Berkeley, CA 94609
}%
\affiliation{%
Kavli Energy NanoScience Institute, Berkeley, CA 94609
}%
\affiliation{%
Materials Science Division, Lawerence Berkeley National Laboratory, Berkeley, CA 94609
}%

\date{\today}
\begin{abstract}
We describe a framework to reduce the computational effort to evaluate large deviation functions of time integrated observables within nonequilibrium steady states. 
We do this by incorporating an auxiliary dynamics into trajectory based Monte Carlo calculations, through a transformation of the system's propagator using an approximate guiding function. 
This procedure importance samples the trajectories that most contribute to the large deviation function, mitigating the exponential complexity of such calculations.
We illustrate the method by studying driven diffusion and interacting lattice models in one and two spatial dimensions.
Our work offers an avenue to calculate large deviation functions for high dimensional systems driven far from equilibrium.
\end{abstract}


\maketitle

Much like their equilibrium counterparts, fluctuations about nonequilibrium steady states encode physical information about a system. 
This is illustrated by the discovery of the fluctuation theorems \cite{Crooks1999,Kurchan1998,lebowitz1999gallavotti,GallavottiCohen1995},  thermodynamic uncertainty relations \cite{barato2015thermodynamic,gingrich2016dissipation}, and  extensions of the fluctuation-dissipation theorem to systems far-from-equilibrium \cite{harada2005equality,speck2006restoring,baiesi2009fluctuations,marconi2008}.
Large deviation functions provide a general mathematical framework within which to characterize and
understand non-equilibrium fluctuations \cite{Touchette2009} and their evaluation has underpinned much recent progress in understanding driven systems \cite{Harris:2007,Mehl:2008dw,Prados:2011,Nyawo:2016}. 
%
%
However,  the current Monte Carlo methods such as the cloning algorithm~\cite{Grassberger2002, delmoral2005, Giardin2006, giardina2011simulating, Pommier2011, Takahiro2016} or transition path sampling \cite{bolhuis2002transition}, exhibit low statistical efficiency when accessing rare fluctuations that are needed to compute them ~\cite{Takahiro2016,ray2017importance}.
This has limited the numerical application of large deviation theory to idealized model systems with relatively few degrees of freedom.


In principle, these difficulties can be eliminated through the use of importance sampling.
A formally exact importance sampling can be derived through 
Doob's h-transform, although this requires the exact eigenvector
of the tilted operator that generates the biased path ensemble~\cite{doobBook, chetrite2015nonequilibrium,sollich2010}.
As this is not practical, approximate importance sampling schemes have been introduced~\cite{klymko2017rare,Takahiro2016,nemoto2017finite}, including a sophisticated iterative algorithm to improve sampling based on feedback and control~\cite{Takahiro2016,nemoto2017finite}.

%

In this Letter, we will show that guiding distribution functions (GDF),  used to implement importance sampling in diffusion Monte Carlo (DMC) calculations of quantum ground states, can be extended to provide an approximate, but improvable,
importance sampling for the simulation of nonequilibrium steady states.
We show the potential of the GDF method
by computing the large deviation functions of time integrated currents at large bias values that capture very rare fluctuations within
two widely-studied models: a driven diffusion model and an interacting lattice model. 
As examples of GDFs, we use analytical expressions as
well GDFs determined from a generalized variational approximation \cite{SI}. The variational approach
provides a procedure to generate guiding functions for arbitrary models of interest.
%

We begin with a short review of the formalism of large deviation functions, drawing
 connections to the ideas of DMC used in this work.
We consider steady states generated by a Markovian dynamics
\begin{equation}\label{Eq:Master}
\partial_t p_t(\mathcal{C})= \W p_t(\mathcal{C})
\end{equation}
where $p_t(\mathcal{C})$ is the probability of a configuration of the system, $\mathcal{C}$, at a time $t$, and $\W$ is a linear operator.
Provided $\W$ is irreducible, it generates a unique steady state in the long time limit that in general produces non-vanishing currents and whose configurations do not necessarily follow a Boltzmann distribution.  
We consider the fluctuations of observables of the form ${\mathcal{O}} = 
 \sum_{t=1}^{t_N} o(\mathcal{C}_{t+},\mathcal{C}_{t-})$, where $o$ is an arbitrary function of configurations at adjacent times, $t+$ and $t-$. 
Within the steady-state, the fluctuations of a time integrated observable can be characterized by a generating function,
\begin{eqnarray}
 \label{Eq:LDF}
 e^{\psi(\lambda) t_N} &=& \left <  e^{-\lambda \mathcal{O}} \right > = \sum_{\mathscr{C}(t_N)} P[\mathscr{C}(t_N)] e^{-\lambda \mathcal{O}} 
\end{eqnarray}
where $\psi(\lambda)$ is the large deviation function,  $\lambda$ is a counting field conjugate to $\mathcal{O}$,
and $P[\mathscr{C}(t_N)] $ is the likelihood of a given trajectory $\mathscr{C}(t_N)=\{\mathcal{C}_0,\mathcal{C}_1,\dots,\mathcal{C}_{t_N}\}$.
%
Derivatives of the large deviation function with respect to $\lambda$ yield the time-intensive cumulants of $\mathcal{O}$. 

In principle, the large deviation function is computable from the largest eigenvalue of a tilted operator $\Wun_\lambda$, i.e., $\Wun_\lambda |\eigv\rangle = \psi(\lambda)  |\eigv\rangle$ where $|\eigv\rangle$ ($\langle\eigv|$) is the corresponding dominant right (left) eigenvector \cite{lebowitz1999gallavotti}.
 In the discreet case, $\Wun_\lambda(\mathcal{C},\mathcal{C}') = \W(\mathcal{C},\mathcal{C}')e^{-\lambda o(\mathcal{C},\mathcal{C}')}(1-\delta_{\mathcal{C},\mathcal{C}'}) -R(\mathcal{C})\delta_{\mathcal{C},\mathcal{C}'}$, where $R(\mathcal{C})=\sum_{\mathcal{C}\neq\mathcal{C}'} \W(\mathcal{C},\mathcal{C}')$ is the exit rate.
For $\lambda = 0$, the tilted operator is Markovian and $\psi(0)=0$ due to normalization.
However, in general $\Wun_\lambda$ does not conserve probability. To sample the dynamics generated
by $\Wun_\lambda$ with Markov chain Monte Carlo, for example, within
the cloning algorithm or transition path sampling,
one must track the normalization with additional weights~\cite{foulkes2001quantum}.
This normalization grows exponentially with $\lambda$. Thus, the associated weights in Monte Carlo algorithms
have an exponentially growing variance, and this is the origin of low statistical efficiency.


The goal of importance sampling is to reduce this variance, and this can be carried out by transforming
the dynamics to restore the normalization of the tilted operator. 
Mathematically, we achieve this through Doob's h-transform,  
\begin{equation}
\W_\lambda(\mathcal{C},\mathcal{C}') =  \eigv(\mathcal{C}) {\Wun}_\lambda(\mathcal{C},\mathcal{C}') \eigv^{-1}(\mathcal{C}') - \psi(\lambda) \, ,
\label{Eq:Doob}
\end{equation}
%
with $\eigv(\mathcal{C}) = \langle \eigv |\mathcal{C}\rangle$. 
We see that $\sum_{\mathcal{C}} \W_\lambda(\mathcal{C},\mathcal{C}') = 0$, since $\sum_{\mathcal{C}} \eigv(\mathcal{C}) {\Wun}_\lambda(\mathcal{C},\mathcal{C}') = \psi(\lambda) \eigv(\mathcal{C}')$. 
The auxiliary dynamics generated by $\W_\lambda$ is thus the optimal dynamics to sample, since the normalization of $\W_\lambda$ is completely independent of configuration. Unfortunately, it requires $\langle\eigv|$ to be known explicitly.  

We can, however, approximate $\langle\eigv|$ and use it to carry out an approximate h-transform
in order to importance sample trajectories. This is the basic technique in this work and is analogous
to using guiding wavefunctions to importance sample
in DMC \cite{Ceperley1980, foulkes2001quantum}. The basic DMC algorithm is equivalent to the cloning algorithm with $\Wun_\lambda$ replaced by the quantum Hamiltonian and  the large deviation function $\psi(\lambda)$ and eigenvector $|\eigv\rangle$ replaced
by the ground-state energy and wavefunction respectively. 
The main technical difference is that unlike the Hamiltonians in DMC, ${\Wun}_\lambda$ is not, in general, Hermitian. 
We focus here on the use of GDFs in the cloning algorithm, though it can also be used with transition path sampling.

%

The cloning algorithm computes the large deviation function as the mixed estimator
\begin{align}
\psi(\lambda) \sim   \frac{1}{t_N} \ln \langle \mathbbm{1}| e^{t_N \Wun_\lambda} |p_0\rangle,
\label{eq:DMC1}
\end{align}
where $\langle \mathbbm{1}| = \sum_\mathcal{C} \langle \mathcal{C}|$ is the uniform left vector, $|p_0\rangle$ is an arbitrary initial state (not orthogonal to the final state) and $\sim$ denotes a long time limit. 
Since the full propagator $\exp[t_N \Wun_\lambda]$ is not known explicitly,  
it is approximated by short-time pieces using a Trotter decomposition, which can be explicitly sampled \cite{foulkes2001quantum}. 
The distribution $p_t$ is represented by an ensemble of walkers, and the propagation $|p_{t+\Delta t}\rangle = \exp [{\Delta t\Wun_\lambda} ] |p_t\rangle$ is obtained via Monte Carlo sampling. As discussed, the sampling procedure accounts for the unnormalized
$\Wun_\lambda$ by keeping weights on the walkers. To avoid a divergence of these weights,
they are redistributed at each iteration, a procedure known as cloning (or branching in DMC)~\cite{Giardin2006, ray2017importance}.
%
%

To importance sample, we now construct an auxiliary dynamics from an approximate GDF, $\langle \app{\eigv} | = \sum_\mathcal{C} \app{\eigv}(\mathcal{C}) \langle \mathcal{C}|$.
We transform Eq.~\ref{eq:DMC1} using the GDF  via the diagonal matrix $\hat{\eigv} = \sum_{\mathcal{C}} \app{\eigv}(\mathcal{C}) |\mathcal{C}\rangle\langle\mathcal{C}|$ such that
\begin{align}
\psi(\lambda) \sim  \frac{1}{t_N} \ln \langle \mathbbm{1}| \hat{\eigv}^{-1} [ \hat{\eigv} e^{t_N \Wun_\lambda}  \hat{\eigv}^{-1}] \hat{\eigv} |p_0\rangle.
\label{eq:DMC2}
\end{align}
The resulting transformed propagator, $\app{W}_\lambda(\mathcal{C},\mathcal{C}') = \app{\eigv}(\mathcal{C}) {\Wun}_\lambda(\mathcal{C},\mathcal{C}') \app{\eigv}^{-1}(\mathcal{C}')$, generates the importance
sampled dynamics.
%
%
Note that $\app{W}_\lambda$ is only Markovian if $\langle \app{\eigv} | = \langle {\eigv}|$, which is generally not the case, and thus the problem of normalization persists. However, if $\langle \app{\eigv} |$ strongly overlaps with $\langle {\eigv}|$, the corresponding exponential growth of the variance is diminished. The key to efficient sampling is thus reduced to determining appropriate approximate GDFs for specific problems. 

We now turn to a numerical assessment of the GDF auxiliary dynamics importance sampling. Here, an important
metric is the efficacy of the auxiliary dynamics. This can be quantified in terms of the statistical efficiency of sampling.
%
For the cloning algorithm, the measure of interest is the number of correlated walkers $\Nc$ \cite{Takahiro2016}. 
%
%
In the case of perfect sampling, using the exact auxiliary dynamics, $\Nc$ is equal to 1. In the other limit, if all walkers are correlated,  $\Nc=\Nw$, the number of walkers used in the simulation.\\

\begin{figure}[!t]
\begin{center}
\includegraphics[width=8.cm]{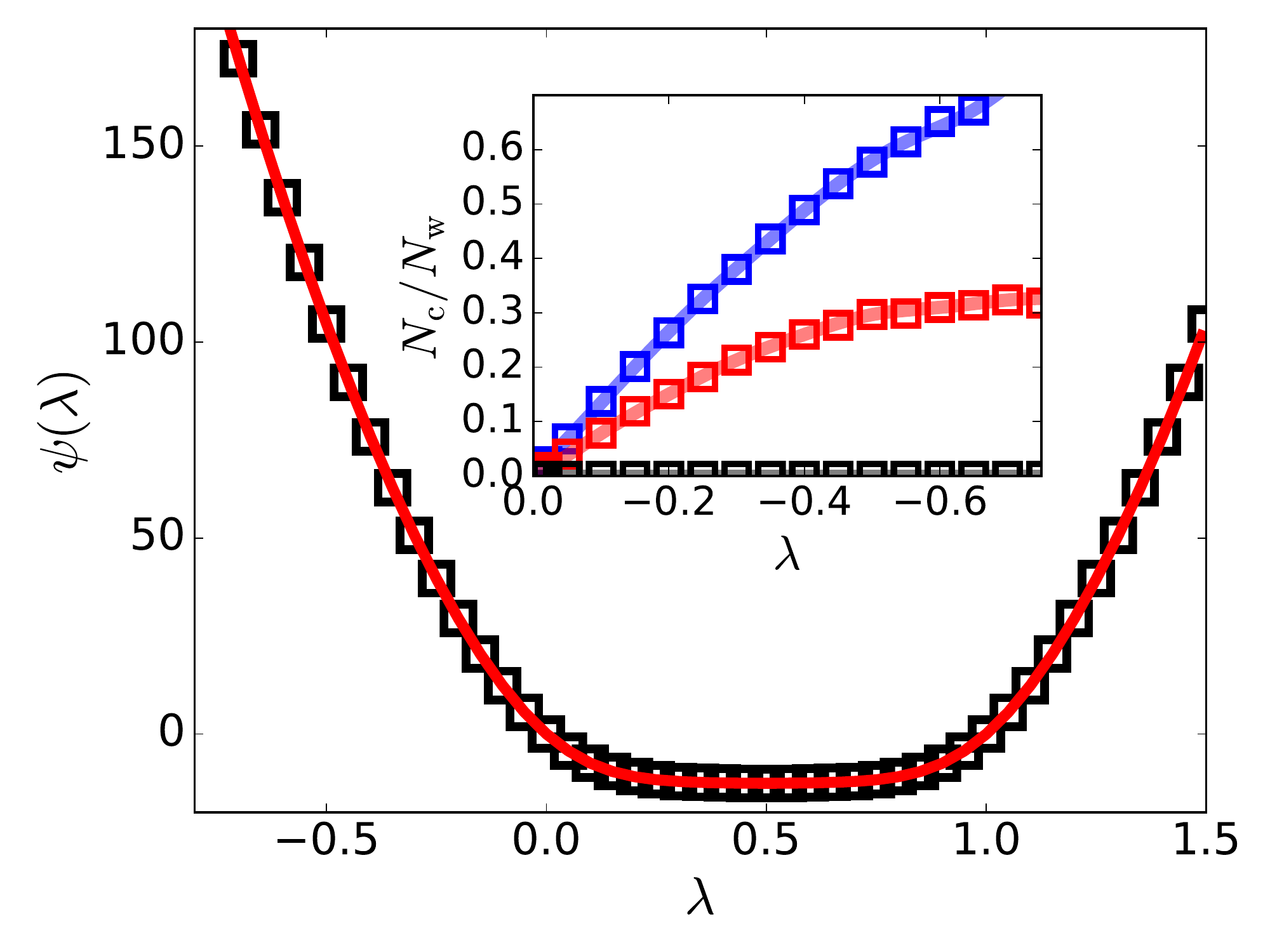}
\caption{Large deviation function for the entropy production of a driven brownian particle on a periodic potential with $v_o=2$, $f=12.5$. The main figure shows the functions computed with exact diagonalization (red) and DMC (black). The inset shows the fraction of correlated walkers without GDF (blue), or with GDF from an instanton approximation to the auxiliary process (red) or the exact auxiliary process (black). }
\label{Fi:1}
\end{center}
\vspace{-20pt} 
\end{figure}

For illustrative purposes, we first consider fluctuations of the entropy production of a driven brownian particle in a periodic potential,
a paradigmatic model in nonequilibrium statistical mechanics.
The equation of motion for the position (on a ring) $\theta$, is $\partial_t{\theta} = F(\theta)+\eta \,$,
with $F(\theta) = f-\partial_\theta V(\theta)$, where $f$ is a constant, nonconservative force, and $V(\theta) = v_o \cos(\theta)$ is a periodic potential \cite{seifert2012stochastic}. The random force, $\eta$, satisfies 
$\langle \eta(t) \rangle=0 \quad$ and $\langle \eta(t)\eta(t') \rangle = 2 \delta(t-t')$.
The entropy production can be computed from $\sigma(\tobs) \tobs = \int_0^{\tobs} f \dot{\theta}(\tau)\mathrm{d}\tau$, which is linearly proportional to the current around the ring \cite{SI}. 

The tilted operator for this model is obtained by absorbing the biasing term $\exp [-\lambda \tobs \sigma(t)]$ into the bare Fokker-Planck propagator, $\W = \partial_\theta^2 - \partial_\theta F(\theta)$, giving
\begin{equation}
{\Wun}_\lambda = \W + 2 f \lambda \partial_\theta  + f\lambda(f\lambda - F(\theta)).
\end{equation}
The last term breaks normalization and must be accommodated through branching. 
The first two terms represent a drift-diffusion process, configurations for which can generated via an associated Langevin equation, 
\begin{equation}
\partial_t{\theta} = F(\theta) - 2f\lambda +\eta.
\label{eq:DBMsde2}
\end{equation}  
Importance sampling this system with a GDF, $\app{\eigv}(\theta)$, produces the transformed propagator
\begin{equation}
\app{W}_\lambda = \W + 2 \partial_\theta[ f\lambda - \partial_\theta \ln \app{\eigv}(\theta)] + \app{\eigv}^{-1}(\theta) {\Wun}_\lambda^\dagger \app{\eigv}(\theta),
\end{equation}
where the adjoint operator is defined as ${\Wun}_\lambda^\dagger = F(\theta)(\partial_\theta-f\lambda) + (\partial_\theta-f\lambda)^2$ \cite{SI}. 
Importance sampled trajectories for $\app{W}_\lambda$ can thus be generated via a Langevin dynamics similar to Eq. \ref{eq:DBMsde2}, but with an additional force $2\partial_\theta \ln \app{\eigv}(\theta)$, and branching weight $\app{\eigv}^{-1}(\theta) {\Wun}_\lambda^\dagger \app{\eigv}(\theta)$.
Note that since the components of the left eigenvector of ${\Wun}_\lambda$ are equal to the components of the right eigenvector of its adjoint ${\Wun}^\dagger_\lambda$, if $\app{\eigv}(\theta)=\eigv(\theta)$, the branching term is equal to $\psi(\lambda)$. 

For this simple one particle system we can determine the optimal GDF by diagonalizing ${\Wun}_\lambda$  in a plane wave basis \cite{Touchette2016, SI}. 
%
  To illustrate the behavior when an approximate GDF, we also consider
a GDF obtained from
an instantonic solution to the eigenvalue equation, which captures the correct limiting behavior of $\eigv(\theta)$ at large $\lambda$, where $\eigv(\theta)$ is just a constant \cite{Touchette2016}.
 
Shown in Fig.~\ref{Fi:1} is the large deviation function computed from exact diagonalization, and cloning algorithm calculations
without a GDF, with the optimal GDF, and with the instantonic GDF. 
All methods converge $\psi(\lambda)$ to good accuracy over the range of $\lambda$, and illustrate the fluctuation theorem symmetry $\psi(\lambda)=\psi(1-\lambda)$. 
However, the statistical effort required to converge the different Monte Carlo calculations varies significantly. 
This is summarized in the inset of Fig.~\ref{Fi:1}, which shows $\Nc$ as a function of $\lambda$. 
The number of correlated walkers increases exponentially without a guiding function, but plateaus if the instantonic guiding function is used.
%
Using the optimal GDF results in walkers that maintain equal weights and stay completely independent, with $\Nc=1$ for all times and all $\lambda$'s.


\begin{figure}[b]
\vspace{-18pt}
\begin{center}
\includegraphics[width=8.cm]{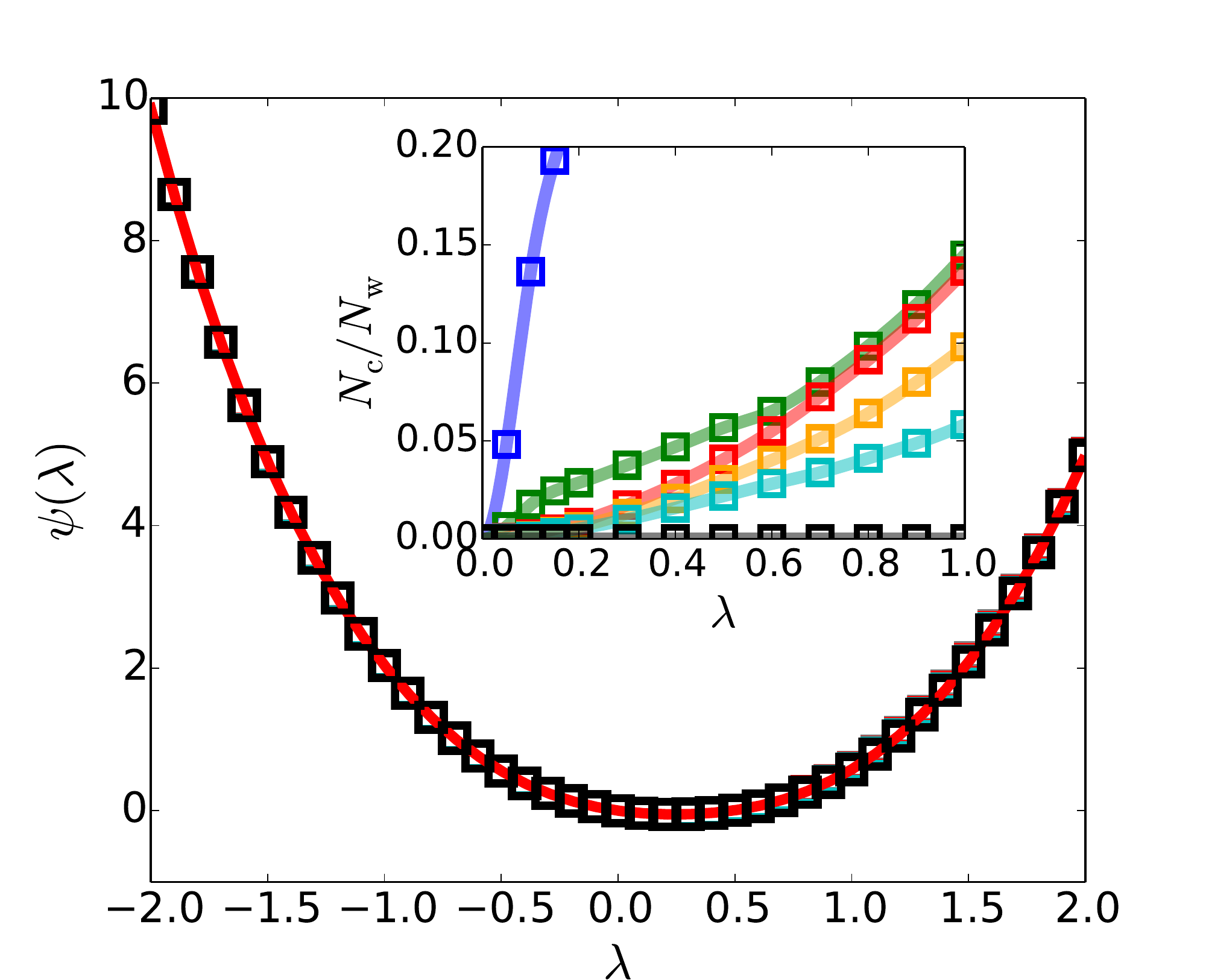}
\caption{Large deviation function for the mass current of an open simple exclusion process. The main figure shows the functions computed with exact diagonalization (red) and DMC (black). The inset shows the fraction of correlated walkers without guiding functions (blue), or with GDF from approximations to the auxiliary process using a uniform GDF (green) or clusters of 1 (red), 2 (orange), 4 (cyan) and 8 (black) sites. }
\label{Fi:2}
\end{center} 
\end{figure}

To explore our framework in different context, we now consider an interacting many-body problem on a lattice, namely the current fluctuations of a simple exclusion process (SEP) \cite{schmittmann1995statistical}. 
The SEP models transport on a lattice with $L$ sites, defined by a set of occupation numbers, $n_i=\{0,1\}$, e.g. $\mathcal{C}=\{0,1,..,1,1\}$. 
%
%
The tilted propagator, $\Wun_\lambda$, has elements corresponding to rates to insert and remove particles at the boundaries if the model is open, with insertion rates $\alpha$ and $\gamma$, and removal rates $\beta$ and $\nu$. 
Within the bulk of the lattice, particles move to the right with rate $p$ and to the left with rate $q$, subject to the constraint of single site occupancy. 
The hard core constraint results in correlations between particles moving on the lattice. 
We consider the large deviation function for mass currents, $Q(\tobs)$, equal to the number of particle hops to the left minus the number of hops to the right, 
\begin{equation}
\label{Eq:Q}
Q(\tobs) = \sum_{t=0}^{\tobs - 1} \sum_{i=0}^{L-1} \delta_{i+1}(t+1)\delta_i(t) - \delta_i(t+1) \delta_{i+1}(t)
\end{equation}
where $\delta_{i}$ is the Kronecker delta function and the sum runs over the lattice site and $\tobs$. 
The propagator is thus dressed by a factor of $\exp [ -\lambda Q(\tobs) ]$. Note that the summand 
is $=0, \pm 1$ depending on particle displacement.  

\begin{figure}[t]
\begin{center}
\includegraphics[width=8.5cm]{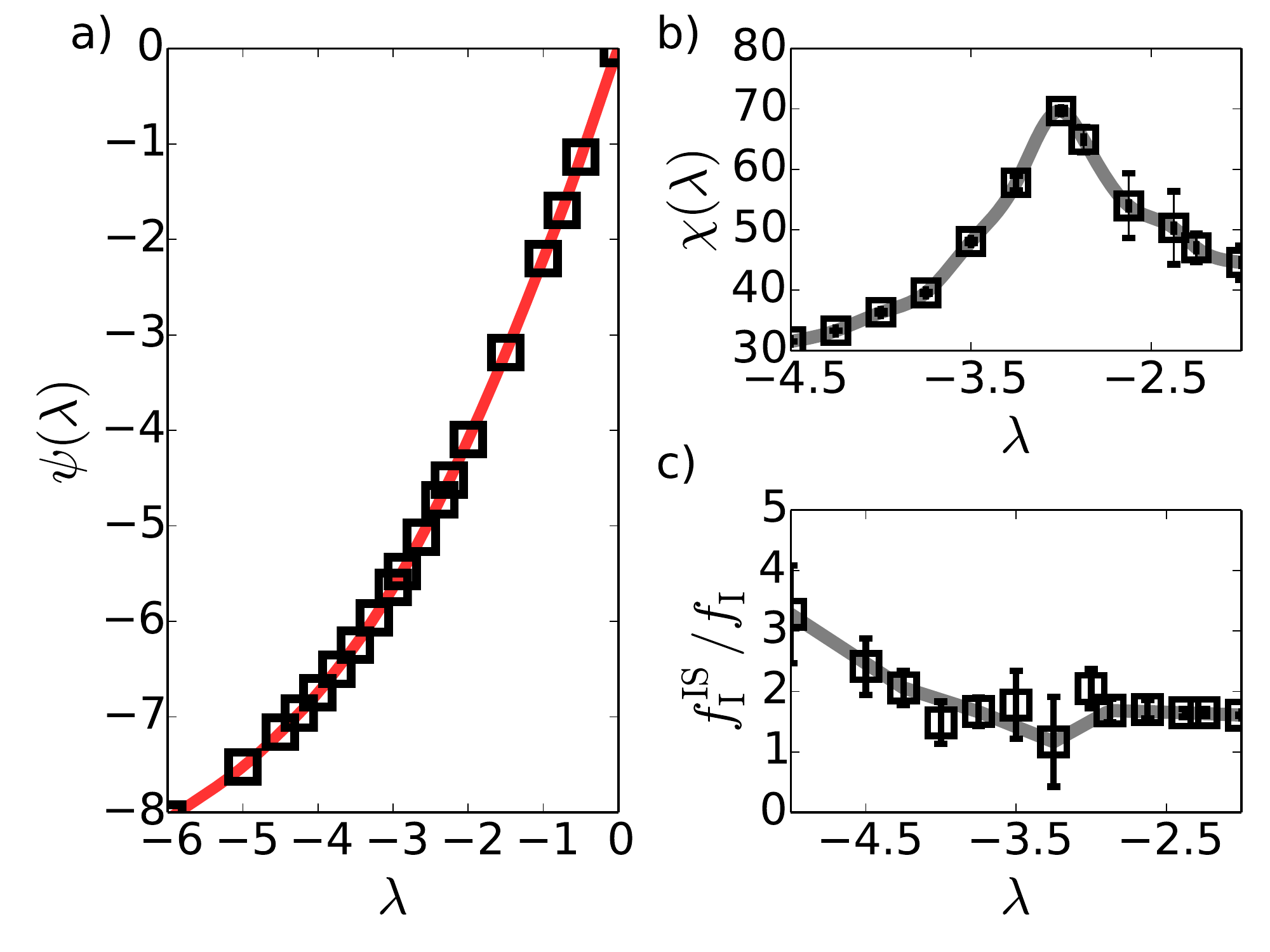}
\caption{Large deviation function for the mass current of a closed 2$d$ asymmetric exclusion process. a) Large deviation function computed from DMC with importance sampling. b) Susceptibility for current fluctuations as a function of $\lambda$. c) Ratio of the fraction of independent walkers with importance sampling, $f_\mathrm{I}^{IS}$, and without importance sampling, $f_\mathrm{I}$.}
\label{Fi:3}
\end{center} 
\vspace{-20pt}
\end{figure}

For all but the smallest lattices, direct diagonalization of ${\Wun}_\lambda$ is impossible, as the size of the matrix scales exponentially with $L$. 
However, we can find an approximate set of eigenvectors using a cluster based mean-field approximation \cite{SI}. 
For example, we can write $|\app{\eigv}\rangle$ as product state of single sites expanded in a basis of single particle states,
$|\app{\eigv} \rangle = \prod_{i=1}^L \sum_{n=0,1} \xi_i(n) | n_i \rangle 
$
where $\xi_i(n)$ are the site expansion coefficients. These can be obtained numerically from the mean-field equations through
a generalized variational principle since ${\Wun}_\lambda$ is not Hermitian (SI), where the stationary solution is found through self-consistent iteration.
%
Similarly, one can
consider a product state of clusters of sites, or a cluster mean-field. Increasing the size of the cluster systematically improves the GDF.

Figure \ref{Fi:2} shows the results of using the cluster mean-field ansatz as the GDF for clusters of different size.
We find that for an $L=8$ lattice, symmetric SEP model \footnote{The model we study has $p = q = 0.5$, $\alpha=\beta=0.9$ and $\gamma=\nu=0.1$.}, all cloning calculations agree with the numerically exact result, and again illustrate fluctuation theorem symmetry, with $\psi$ symmetric about half of the current's affinity \cite{lebowitz1999gallavotti}.
The statistical effort needed to converge each calculation is decreased by several orders of magnitude when a GDF is used. 
As shown in the inset of Fig. \ref{Fi:2}, even with an auxiliary dynamics computed from the single-site mean field theory, the fraction of independent walkers, $f_\mathrm{I}=1-\Nc/\Nw$, is increased by a factor of 40, and this efficiency is systematically improved with auxiliary dynamics computed from larger cluster states. 
As before, if the exact auxiliary process is used, corresponding to a cluster of 8 sites, $\Nc=1$ for all times and all $\lambda$'s.

As a third illustration of the auxiliary dynamics framework, we consider a 2D generalization of a closed SEP model in the presence of a weak external field that biases transport in one direction \cite{SI}. 
This system has been considered recently \cite{tizon2016order}, where it was found that its large deviation function for current fluctuations in the direction of the driving exhibits a dynamical phase transition. 
For small $\lambda$, the system is in a homogeneous phase, while for large negative $\lambda$ the system phase separates, forming a traveling wave in the direction of the biased current. 
We find critical behavior for a 12$\times$12 lattice as illustrated in Figs. 3a,b, where for $\lambda \approx-3$ the fluctuations in the current, $\chi(\lambda)=d^2 \psi/d\lambda^2$, are maximized and presumed to diverge in the infinite system limit \footnote{For the 2D SEP model, we study a 12$\times$12 lattice, with hopping rates in the $x$ and $y$ direction that are now vectorial with components $\mathbf{p}=\{0.22,0.5\}$ and $\mathbf{q}=\{1.15,0.5\}$ and bias on the current in just the $x$ direction as in Eq. ~\ref{Eq:Q}.}.
Beyond this critical value, the state of the system is fluctuation dominated, and as such serves as a good test of our importance sampling methodology. 
Shown in Fig.~\ref{Fi:3}c is the ratio of the fraction of independent walkers, $f_\mathrm{I}$, computed using a 4$\times$2 sites cluster mean field GDF and without importance sampling, as a function of $\lambda$.
For small $|\lambda|$ the bare dynamics is capable of sampling the biased distribution and the enhancement from importance sampling is $\sim 2$. 
However, even for the traveling wave state, where the system is not well described by mean field theory, we find an increased sampling efficiency by a factor of 2-4 over bare sampling \cite{SI}. This result shows that even a poor approximation to the steady-state using the cluster approach is able to aid the convergence of $\psi(\lambda)$.\\ 
%

Beyond these specific examples, we emphasize that the guiding framework we have described is general and is not restricted to the models we have considered.
 For example, we can consider importance sampling an $N$-particle interacting continuum dynamics generated by a Fokker-Planck operator $\W$ for an arbitrary many-body force $\mathbf{F}_i({\bf r}_1,\ldots,{\bf r}_N)$ in $d$ dimensions. 
The GDF in this case is an $N$-particle function $\app{\eigv}(\mathbf{r}_1,\ldots, \mathbf{r}_N)$ and the transformed tilted operator for the large deviation function for the total mass current vector $\mathbf{J}(\tobs) =\sum_i \int_0^{\tobs} \dot{\mathbf{r}}_i(\tau) d\tau$, 
\begin{align}
\tilde{W}_{\bm{\lambda}} = &\sum_i \nabla_i^2 - \nabla_i \cdot \left [ \mathbf{F}_i({\bf r}_1,\ldots,{\bf r}_N) - 2\bm{\lambda} + 2\nabla_i \ln {\app{\eigv}} \right ] \nonumber \\ 
&+ \app{\eigv}^{-1} \mathbb{W}_\lambda^\dagger \app{\eigv}
\end{align}
where the first two terms are the drift-diffusion terms, and the last  term gives the branching weight, which is $\mathbb{W}^\dagger_{\bm{\lambda}} =  \sum_i \nabla_i^2 + (\mathbf{F}_i-2\bm{\lambda})\cdot \nabla_i 
+ \bm{\lambda}\cdot (\bm{\lambda} - \mathbf{F}_i) $. Here, $\bm{\lambda}$ is a $d$-dimensional vector, biasing the independently different components of the current.
Approximating $\app{\eigv}$ can then be done by choosing a
  $N$-particle functional form, whose parameters are determined by a generalized variational procedure similar to that used in the determination of the
  cluster mean-field GDF above. 
This extends what is done in DMC, where guiding functions are first determined by a variational Monte Carlo procedure. 
As a simple choice in the continuum, one could use a product state
$\app{\eigv}= f_1(\mathbf{r}_1)   f_2(\mathbf{r}_2)\ldots f_N(\mathbf{r}_N)$, or a product of pairs introduced with Jastrow factors $\app{\eigv}=\prod_{i<j} \exp [\mathcal{J}(\mathbf{r}_i,\mathbf{r}_j)]$. In the lattice
setting matrix product or tensor network states of low bond dimension appear as natural guiding functions~\cite{Evans:2007,Chan:2014}.
The use of such numerically determined GDFs ensures that the importance sampling  captures the general influence of the interactions generated at a given bias, $\lambda$.


In conclusion, the use of guiding functions to importance sample the trajectory space of nonequilibrium steady states
makes computing large deviation functions possible in complex systems. The formalism we have used is
applicable to any non-equilibrium state generated by a deterministic master equation, while the
variational determination of the guiding function provides a systematic way to importance sample
non-equilibrium problems for which analytical information on the solution is not known. 
These techniques open up the possibility to study ever larger systems, for longer times, with increased molecular resolution. 


{\bf Acknowledgements:} The authors would like to thank Rob Jack, Vivien Lecomte, Juan P. Garrahan and David Ceperley for fruitful and engaging discussions. D.T.L was supported by UC Berkeley College of Chemistry. U. R. was supported by the Simons Collaboration on the Many-Electron Problem and the California Institute of Technology. G. K.-L. C. is a Simons Investigator in Theoretical Physics and was supported by the California Institute of Technology and the US Department of Energy, Office of Science via DE-SC0018140.  These calculations were performed with CANSS, available at https://github.com/ushnishray/CANSS.

\bibliography{v1}
\end{document}


\title{Exact fluctuations of nonequilibrium steady states \\from approximate auxiliary dynamics:\\Supplementary Information}
\author{Ushnish Ray}
\affiliation{%
Division of Chemistry and Chemical Engineering, California Institute of Technology, Pasadena, CA 91125
}%
\author{Garnet Kin-Lic Chan}
\affiliation{%
Division of Chemistry and Chemical Engineering, California Institute of Technology, Pasadena, CA 91125
}%
\author{David T. Limmer}
\affiliation{%
Department of Chemistry, University of California, Berkeley, CA 94609
}%
\affiliation{%
Kavli Energy NanoScience Institute, Berkeley, CA 94609
}%
\affiliation{%
Materials Science Division, Lawerence Berkeley National Laboratory, Berkeley, CA 94609
}%

\date{\today}
\pacs{}
\keywords{} 
\maketitle

\section{Guiding distribution function in the continuum}

Here we derive the tilted propagator used for importance sampling the driven brownian walker in the main text and discuss the adjoint operator for generating the Langevin dynamics. The Langevin equation that governs the overdamped motion of a particle in the presence of an external force, $F(\theta)$, is given by,
\begin{align}
\partial_t\theta = F(\theta) + \eta
\end{align}
As in the main text, the associated Fokker-Planck equation corresponding to this Langevin equation is
\begin{equation}
\partial_t p_t(\theta) = \mathcal{W}p_t(\theta)
\end{equation}
where $p_t(\theta)$ is the probability of observing the particle at $\theta$ at a time $t$ and
\begin{align}
\label{eq:supp_dbm2}
\W = -\partial_\theta(F-\partial_\theta)
\end{align} 
is the Fokker-Planck operator.
We are interested in the large deviation function of the entropy production, $s(\tobs)=\sigma(\tobs)\tobs$, which is proportional to the current around the ring, unwrapped so that winding numbers are included,
\begin{align}
s(\tobs) = \int^{t_N}_0 f \dot{\theta}(\tau) d\tau = f\Delta\theta(\tobs) \equiv fx
\end{align}
The generating function is related to the probability $p(\theta,s,\tobs)$ of finding the particle at $\theta$ with entropy produced $s$ at a time $\tobs$ by the Laplace transform,
\begin{align}
\rho(\theta,\lambda,t) = \int ds \, e^{-\lambda s} p(\theta,s,\tobs).
\label{eq:supp_dbm1}
\end{align}
Since we are interested in the behavior of probability distribution conditioned on $x$ and $\theta$ and since both of them share the same noise source $\eta$, we expand the Fokker-Planck operator via 
$\partial_\x \rightarrow \partial_\x + \partial_x$ to obtain:
 \begin{align}
 \tilde{\W} = \W + (2\partial_\x - F(\x))\partial_x + \partial^2_x
 \end{align}
and the corresponding modified Fokker-Planck equation is
\begin{align}
\partial_t p(\x,s,t) =  \tilde{\W}p(\x,s,t).
\label{eq:supp_dbm3}
\end{align}
By differentiating Eq. \ref{eq:supp_dbm1} with respect to $t$ and inserting Eq. \ref{eq:supp_dbm3} we get
\begin{align}
\partial_t\rho(\x,\lambda,t) = \int ds e^{-\lambda s} \tilde{\W} p(\x,s,t). 
\end{align}
Now,
\begin{align}
 \int ds e^{-\lambda s} \tilde{\W} p(\x,s,t)  =  &\int ds e^{-\lambda s} \W p(\x,s,t) \nonumber\\
 + &\int ds e^{-\lambda s}  (2\partial_\x - F(\x))\partial_x p(\x,s,t) \nonumber\\
 + &\int ds e^{-\lambda s}  \partial^2_x p(\x,s,t)
 \end{align}
Performing integration by parts we get:
\begin{align}
\partial_t \rho(\x,\lambda,t)  =  &\W \int ds e^{-\lambda s} p(\x,s,t) \nonumber\\
 +  &(2\partial_\x - F(\x))(f\lambda) \int ds e^{-\lambda s} p(\x,s,t) \nonumber\\
 +  &(f\lambda)^2 \int ds e^{-\lambda s} p(\x,s,t) \nonumber\\
= &\Wun_\lambda \int ds e^{-\lambda s} p(\x,s,t) = \Wun_\lambda \rho(\x,\lambda,t),
\end{align}
where, 
\begin{align}\
\label{Eq:Tilted}
{\Wun}_\lambda = \W + 2 f \lambda \partial_\theta  + f\lambda(f\lambda - F(\theta)),
\end{align}
is the tilted operator given by Eq. 8 in the main text used to obtain the modified or tilted dynamics \cite{Mehl:2008dw}. The mapping of the second-order differential operator
onto a stochastic diffusion process then follows the well-known Feynman-Kac theorem \cite{chetrite2015nonequilibrium}. 

Following the general derivation of importance sampling in the main text, we use the specific form of $\Wun_\lambda$ to get
\begin{align}
\tilde{\Xi} \Wun_\lambda (\tilde{\Xi}^{-1}\rho) = \rho[\tilde{\Xi} \Wun_\lambda \tilde{\Xi}^{-1} + \partial_\theta \zeta] - \partial_\theta(\zeta - \partial_\theta)\rho,
\end{align}  
with $\zeta = F - 2f\lambda + 2\partial_\theta\ln\tilde{\Xi}$. 
The last term corresponds to a Fokker-Planck type of operator with a force term equal to $\zeta$ and can be used to generate trajectories. The first term changes normalization and is, therefore, used for branching. The equation of motion for $\theta$ is obtained from the adjoint operator $\mathcal{W}_\lambda^\dagger$, where,
\begin{align}
\tilde{\Xi}^{-1}\Wun_\lambda^\dagger\tilde{\Xi} = \tilde{\Xi} \Wun_\lambda \tilde{\Xi}^{-1} + \partial_\theta \zeta
\end{align}
which follows from integration by parts. The specific form of the branching term following the operators in the main text is given by 
\begin{align}
\tilde{\Xi}^{-1}\Wun_\lambda^\dagger\tilde{\Xi} = \frac{1}{\tilde{\Xi}}\frac{d^2\tilde{\Xi}}{d\theta^2} + (F-2\lambda f)\frac{d \ln \tilde{\Xi}}{d\theta} + f\lambda (f\lambda - F(\theta)),
\end{align}
where $F(\theta) = f-\partial_\theta V(\theta)$, with $f$ a constant, nonconservative force, and $V(\theta) = v_o \cos(\theta)$ is a periodic potential. 

As discussed in the text, the exact GDF can be computed by expanding $\tilde{\Xi}$ is plane waves. Specifically, we express 
\begin{align}
\tilde{\Xi}(\theta) = \sum_{n=-m}^{m} c_n e^{i n \theta}
\end{align}
and inserting this expansion into eigenvalue relation, and using the tilter operator in Eq.~\ref{Eq:Tilted}, we can derive a recursion relationship for the coefficients $c_n$. This yields,
\begin{eqnarray}
c_n a_n - c_{n-1} b_{-}  - c_{n+1} b_{+}  = c_n \psi(\lambda)
\end{eqnarray}
with elements
$$
a_n =  i n (f+f\lambda )-n^2/2+\lambda f^2+\lambda^2 f^2/2
$$
$$
b_{\pm} = \pm v_o(-i\lambda f+n \pm 1)/2
$$
which specifies a tridiagonal matrix that can be diagonalized using standard techniques suitable for non-Hermitian matrices. In practice, we use $2M+1$ basis functions with $M = 50$ for the exact calculations. The instantonic solution is generated using only 1 basis function. Note that Ref. \cite{Touchette2016} contains an error in the typesetting of the equations defining the expansion coefficients. 

\section{Guiding distribution functions from discrete mean-field solutions}

In this section we outline the procedure needed to generate mean-field (MF) and cluster solutions that
form the guiding distribution functions in our discrete models. We will illustrate the procedure for the SEP but it can be easily generalized to other models.

For the open boundary SEP the tilted matrix may be written as:
\begin{align}
\Wun_\lambda = {w}_L + \sum_{i=1}^{L} {w}_i + {w}_R
\end{align}
where the single particle transition matrices  $\{{w}_L, {w}_i, {w}_R\}$  and their operator representations are given by
\begin{align}
{w}_L &= \left[ 
\begin{array}{cc}
-\alpha & \gamma e^{\lambda} \\
\alpha e^{-\lambda} & -\gamma \\
\end{array}
\right]  
\quad \quad
{w}_R &= \left[ 
\begin{array}{cc}
-\nu & \beta e^{-\lambda} \\
\nu e^{\lambda} & -\beta \\
\end{array}
\right]
\quad \quad
\end{align}
and
\begin{align}
{w}_i &= \left[ 
\begin{array}{cccc}
 0 & 0 & 0 & 0\\
 0 & -q & p e^{-\lambda} & 0\\
 0 & qe^{\lambda} & -p & 0\\ 
 0 & 0 & 0 & 0 
\end{array} 
\right]
%
\end{align}
 
These matrices can be combined to form a many-body matrix that in operator notation is written as\begin{widetext}
\begin{align}
\Wun_\lambda = &-\alpha(\mathbbm{1}-\hat{n}_1) + \alpha e^{-\lambda} \hat{c}_1^\dagger + \gamma e^{\lambda}\hat{c}_1 -\gamma \hat{n}_1 + pe^{-\lambda}\hat{c}^\dagger_2\hat{c}_1 - p\hat{n}_1(\mathbbm{1}-\hat{n}_2) \nonumber\\
&+ \sum_{i=2}^{L-1} pe^{-\lambda} \hat{c}^\dagger_{i+1}\hat{c}_i - p\hat{n}_i(\mathbbm{1}-\hat{n}_{i+1}) + qe^{\lambda}\hat{c}^\dagger_{i-1}\hat{c}_i - q\hat{n}_i(\mathbbm{1}-\hat{n}_{i-1}) \nonumber\\
&-\nu(\mathbbm{1}-\hat{n}_L) + \nu e^{\lambda} \hat{c}_L^\dagger + \beta e^{-\lambda}\hat{c}_L -\beta \hat{n}_L + qe^{\lambda}\hat{c}^\dagger_{L-1}\hat{c}_L - q\hat{n}_L(\mathbbm{1}-\hat{n}_{L-1}).
\label{eq:SEPM}
\end{align}
\end{widetext}
Here $\hat{c}^\dagger_i$ creates a particle at site $i$, $\hat{c}_j$ destroys a particle at site $j$, and $\hat{n}_i$ counts the number of particles at site $i$. The combined operators $\hat{c}^\dagger_i\hat{c}_j$ correspond to kinetic-like terms that move a particle from site $j$ to $i$, and $\hat{n}_i(\mathbbm{1}-\hat{n}_j)$ represents a hard-core interaction that prevents double occupation of sites. The exact solution is the usual eigenvalue problem ${\Wun_\lambda}|\Xi\rangle = \varepsilon|\Xi\rangle$ where $(\varepsilon,|\Xi\rangle)$ is a particular eigenpair (the inverse eigenvectors $\{\langle \Xi|\}$ can be constructed from $\{|\Xi\rangle\}$).
 
An obvious route to explore in constructing approximate GDF is to use a mean-field (MF) solution. The MF approach is to
approximate the many-body state as a product state. Starting with a product of single site states, we can systematically improve our results by moving to products of cluster states, where interactions in the cluster are treated explicitly, while the links between clusters are treated at the MF level. We first illustrate the procedure for single site clusters and then show how to
generalize to multi-site clusters.
\vspace{-10pt}

\subsection{Site-Decoupled Mean-Field and Generalized Variational Approximation}

For the single site clusters we approximate the solution by the form $|\tilde{\Xi}\rangle = \prod_{i=1}^L |\xi_i\rangle$, where the
single site state $|\xi_i\rangle = \sum_{n=0}^1 \xi_i (n) |n_i\rangle$ is written in a basis of occupation numbers $\{|n\rangle\}$. Here $\xi_i(n)$ is a scalar function dependent on the occupation number $n$. One subtlety is that because the tilted matrix is not Hermitian
we must distinguish between approximating its left and its right eigenvector. In this work, we will always approximate
the right eigenvector. The corresponding left eigenvector is then made of the product of left states
$\langle \xi_i| = \sum_{n=0}^1 {\xi}'_i(n) \langle n_i |$, where $\langle \xi_i|$ (and the coefficients $\xi_i'$) can be obtained
from the right states by the biorthogonality condition, $\langle \xi_i|\xi_j \rangle = \delta_{ij}$.
For the following discussion $\{|\tilde{\Xi}\rangle, \langle\tilde{\Xi}|\}$ represents the biorthogonal pair.


To determine the coefficients $\xi_i(n)$, we will use a generalized variational procedure to make the
functional
\begin{align}
\langle\tilde{\Xi}| (\Wun_\lambda - \varepsilon)|\tilde{\Xi}\rangle \label{eq:bifunctional}
\end{align}
stationary with respect to independent linear
variations in $\langle \tilde{\Xi}|$ and $|\tilde{\Xi\rangle}$. Although stationarity of Eq.~\ref{eq:bifunctional}
does not give a minimum principle, it leads to a mean-field approximation that approximates the right (or left) eigenvector
equation when projected into an appropriate space, similar to the use of collocation to solve a partial differential equation.
For example, considering stationarity with respect to small variations in the left state, we obtain
\begin{align}
  \langle \delta\tilde{\Xi} | \Wun_\lambda - \varepsilon | \tilde{\Xi}\rangle = 0 \label{eq:varfunc}
  \end{align}
The small variations in the left state take the explicit form
\begin{align}
\langle \delta\tilde{\Xi}| &= 
\sum_{i=1}^L \langle \delta \xi_i| \prod_{j\ne i}^L \langle \xi_j| \nonumber\\  
&= \sum_{i=1}^L \left[ \sum_{n_i=0}^1 \delta \xi'_i(n)\langle n_i| \right] \prod_{j\ne i}^L\left[\sum_{n_j=0}^1 \xi'_i(n)\langle n_j|\right] \label{eq:varstate}
\end{align}
We then evaluate Eq.~\ref{eq:varfunc} with respect to the specific variations in Eq.~\ref{eq:varstate} to obtain
mean-field equations.
Writing $\Wun_\lambda = ({\mathcal{W}}_\lambda-{V}_i) + {V}_i$ where ${V}_i$ represents all terms in ${\mathcal{W}}_\lambda$ that contains terms involving site $i$, we obtain the following equation for each $i$:
\begin{align}
\langle \delta \tilde{\Xi}| {V}_i |\tilde{\Xi}\rangle -\epsilon_i \langle\delta {\xi}_i| {\xi}_i\rangle = 0, 
\label{eq:mfvareq}
\end{align}
which is equivalent to an eigenvalue equation with eigenvalue $\epsilon_i$, 
where we have defined
\begin{align}
\epsilon_i &= \varepsilon - \left[\prod_{j\ne i}\langle {\xi}_j|\right] 
(\Wun_\lambda - {V}_i) 
\left[\prod_{k\ne i} | {\xi}_k\rangle\right] 
\end{align}
As we are interested in the  maximal eigenvalue/eigenvector pair of $\Wun_\lambda$, we are interested
in the maximal eigenvalue/eigenvector pair for each site $i$.

For the SEP $\Wun_\lambda$, Eq. \ref{eq:mfvareq} implies solving the following decoupled eigenvalue equation  for $i = 1$
\begin{align}
&[(-\alpha-q\langle\hat{n}_2\rangle)(\mathbbm{1}-\hat{n}_1) 
+ (pe^{-\lambda}\langle\hat{c}^\dagger_2\rangle + \gamma e^{\lambda})\hat{c}_1\nonumber\\
&+ (\alpha e^{-\lambda}  + qe^{\lambda}\langle\hat{c}_2\rangle))\hat{c}_1^\dagger
+ (-\gamma-p(1-\langle\hat{n}_2\rangle))\hat{n}_1] |\xi_1\rangle \nonumber\\
&= \varepsilon_1|\xi_1\rangle
\end{align}
where we have used the shorthand $\langle \ldots \rangle$ to denote the expectation value
$\langle \tilde{\Xi} | \ldots | \tilde{\Xi}\rangle$.
For $i = L$, the eigenvalue equation is
\begin{align}
&[(-\nu-p\langle\hat{n}_{L-1}\rangle)(\mathbbm{1}-\hat{n}_L) 
+ (qe^{\lambda}\langle\hat{c}^\dagger_{L-1}\rangle + \beta e^{-\lambda})\hat{c}_{L-1}\nonumber\\
&+ (\nu e^{\lambda}  + pe^{-\lambda}\langle\hat{c}_{L-1}\rangle))\hat{c}_L^\dagger
+ (-\beta-q(1-\langle\hat{n}_{L-1}\rangle))\hat{n}_L] |\xi_L\rangle \nonumber\\
&= \varepsilon_L|\xi_L\rangle
\end{align}
Finally for all other $i$, the eigenvalue equation is
\begin{align}
&[(-p\langle\hat{n}_{i-1}\rangle - q\langle\hat{n}_{i+1}\rangle)(\mathbbm{1}-\hat{n}_i) \nonumber\\
&+(pe^{-\lambda}\langle\hat{c}^\dagger_{i+1}\rangle + qe^{\lambda}\langle\hat{c}^\dagger_{i-1}\rangle)\hat{c}_{i}\nonumber\\
&+ (qe^{\lambda}\langle\hat{c}_{i+1}\rangle + pe^{-\lambda}\langle\hat{c}_{i-1}\rangle)\hat{c}_{i}^\dagger \nonumber\\
&+ (-p(1-\langle\hat{n}_{i+1}\rangle)-q(1-\langle\hat{n}_{i-1}\rangle))\hat{n}_i] |\xi_i\rangle \nonumber\\
&= \varepsilon_i|\xi_i\rangle \, .
\end{align}

The maximal eigenvalue/eigenvector pair approximation can be obtained by solving the above equations for each site $i$ and choosing the set of states $\{|\xi_i\rangle\}$ corresponding to the largest eigenvalues $\{\epsilon_i\}$. Notice that even when collecting terms involving site $i$, there will inevitably be terms involving other sites due to the two-body interaction present in the matrix (\ref{eq:SEPM}). The natural way to solve this system, thus, requires the use of a self-consistent procedure. We start with an initial set of guess values for $\langle \hat{n}_i\rangle$, $\langle \hat{c}_i\rangle$ and $\langle \hat{c}^\dagger_i\rangle$ and proceed to solve the individual eigenvalue problems for each site. At the end of each iteration we use the states $\{|\xi_i\rangle\}$ to recompute $\langle \hat{n}_i\rangle$, $\langle \hat{c}_i\rangle$, and  $\langle \hat{c}^\dagger_i\rangle$ and use them for the next iteration. This is continued until $\{\langle \hat{n}_i\rangle, \langle \hat{c}_i\rangle, \langle \hat{c}^\dagger_i\rangle\}$ do not change.\\

Once the solutions have converged it is straightforward to obtain the MF estimate of the LDF, $\epsilon(\lambda) = \langle \tilde{\Xi} | \Wun_\lambda | \tilde{\Xi} \rangle$. The MF approximation to the state $\langle\tilde{\Xi}|$ is precisely the GDF that we need to construct the auxiliary process that will importance sample the LDF. The effective matrices incorporating the importance sampling are:

\begin{align}
\tilde{w}_L &= \left[ 
\begin{array}{cc}
-\alpha & \gamma e^{\lambda} \frac{\xi_1(0)}{\xi_1(1)}\\
\alpha e^{-\lambda} \frac{\xi_1(1)}{\xi_1(0)} & -\gamma \\
\end{array}
\right]  
\end{align}

\begin{align}
\tilde{w}_i &= \left[ 
\begin{array}{cccc}
 0 & 0 & 0 & 0\\
 0 & -q & p e^{-\lambda} \frac{\xi_i(0) \xi_{i+1}(1)}{\xi_i(1) \xi_{i+1}(0)}& 0\\
 0 & qe^{\lambda} \frac{\xi_{i-1}(1) \xi_i(0)}{\xi_{i-1}(0) \xi_i(1)} & -p & 0\\ 
 0 & 0 & 0 & 0 
\end{array} 
\right]
\end{align}

\begin{align}
\tilde{w}_R &= \left[ 
\begin{array}{cc}
-\nu & \beta e^{-\lambda} \frac{\xi_{L}(0)}{\xi_{L}(1)}\\
\nu e^{\lambda} \frac{\xi_{L}(1)}{\xi_{L}(0)} & -\beta \\
\end{array}
\right]
\end{align}
Notice that these are not normalized and their renormalization determines the branching weights. 
In the text, we use a first order Trotterization on the short-time importance-sampled propagator ($\langle \tilde{\Xi}| e^{dt\Wun_\lambda} |\tilde{\Xi}\rangle$) to obtain the Markovian transition probability matrix $\tilde{\mathbb{U}}_\lambda \equiv \mathbbm{1} + dt\langle \tilde{\Xi}| \Wun_\lambda  |\tilde{\Xi}\rangle$.
Therefore, $\tilde{\mathbb{U}}_\lambda$ follows directly from the transformed components of $\Wun_
\lambda$, and is given by
\begin{align}
\tilde{\mathbb{U}}_\lambda = \tilde{u}_L + \sum_{i=1}^{L} \tilde{u}_i + \tilde{u}_R,
\end{align}
where,
\begin{align}
\tilde{u}_L &= \left[ 
\begin{array}{cc}
1-\alpha dt & \gamma e^{\lambda} \frac{\xi_1(0)}{\xi_1(1)}\\
\alpha e^{-\lambda} \frac{\xi_1(1)}{\xi_1(0)} & 1-\gamma dt \\
\end{array}
\right]
\label{su:eq1}  
\end{align}

\begin{align}
\tilde{u}_i &= \left[ 
\begin{array}{cccc}
1 & 0 & 0 & 0\\
 0 & 1-qdt & p e^{-\lambda} \frac{\xi_i(0) \xi_{i+1}(1)}{\xi_i(1) \xi_{i+1}(0)}& 0\\
 0 & qe^{\lambda} \frac{\xi_{i-1}(1) \xi_i(0)}{\xi_{i-1}(0) \xi_i(1)} & 1-pdt & 0\\ 
 0 & 0 & 0 & 1 
\end{array} 
\right]
\label{su:eq2}
\end{align}

\begin{align}
\tilde{u}_R &= \left[ 
\begin{array}{cc}
1-\nu dt & \beta e^{-\lambda} \frac{\xi_{L}(0)}{\xi_{L}(1)}\\
\nu e^{\lambda} \frac{\xi_{L}(1)}{\xi_{L}(0)} & 1-\beta dt \\
\end{array}
\right]
\label{su:eq3}
\end{align}
are the associated transition probabilities at the ends of the lattice, $\tilde{u}_L$ and $\tilde{u}_R$, or in its interior, $\tilde{u}_i$. 
At every time step or Monte Carlo sweep, the current state $|\mathcal{C}\rangle = |n_1 n_2 ... n_L\rangle$ that corresponds to a column of $\tilde{\mathbb{U}}_\lambda$ is used to propose moves such that the outgoing state $|\mathcal{C}'\rangle$ has the probability $\tilde{\mathbb{U}}_\lambda(\mathcal{C}',\mathcal{C}) /\mathcal{N}(C)$ of being accepted, where $\mathcal{N}(C) \equiv \sum_{\mathcal{C'}} \tilde{\mathbb{U}}_\lambda(\mathcal{C}',\mathcal{C})$ is the normalization factor. Over the course of the short trajectory generated in between branching steps, a walker's weight is accumulated as a product of these normalization factors.\\ 

\subsection{Cluster Approach}
The idea of a site-decoupled mean-field can be extended to multiple sites collected into clusters. The MF ansatz is
then $|\tilde{\Xi}\rangle = \prod^{L/c_L}_{c=1} |\xi_c\rangle$ 
where $c_L$ is the number of sites that constitutes a cluster. $|\xi_c\rangle$
is expanded in the occupation basis of the cluster
$|\xi_c\rangle = \sum_{\{n\}} \xi_c(n_{(c-1)c_L+1},...,n_{cc_L})|n_{(c-1)c_L+1},...,n_{cc_L}\rangle$.
For each cluster $(c)$, we can write down the analogous $V_i$ operator
which contains all terms in Eq.~\ref{eq:SEPM} that involve sites in the cluster ($V_c$).
All terms involving only sites inside the given cluster are treated exactly (i.e. treated as operators)
while terms that involve sites that are inside two different clusters are split up via the MF approximation $\hat{A}_i \hat{B}_j \sim \langle \hat{A}_i \rangle \hat{B}_j + \hat{A}_i \langle \hat{B}_j \rangle - \langle \hat{A}_i \rangle \langle \hat{B}_j \rangle$, where as before $\langle \hat{A}_i \rangle$ and $\langle \hat{B}_j \rangle$ are determined self-consistently. 
 
More explicitly, the mean-field cluster $V_c$ operator is
\begin{widetext}
\begin{align}
{V_c} &=  \hat{L}(j = c_L(c-1)+1) + \hat{R}(j = c_Lc-1) \nonumber\\
&+ \sum_{j=c_L(c-1)+1}^{c_Lc-1} pe^{-\lambda} \hat{c}^\dagger_{j+1}\hat{c}_j - p\hat{n}_j(\mathbbm{1}-\hat{n}_{j+1}) + qe^{\lambda}\hat{c}^\dagger_{j}\hat{c}_{j+1} - q\hat{n}_{j+1}(\mathbbm{1}-\hat{n}_{j}) \\
\text{where} \nonumber\\
\hat{L}(j) &= -p\langle \hat{n}_{j-1}\rangle(\mathbbm{1}-\hat{n}_{j}) + qe^{\lambda} \langle \hat{c}^\dagger_{j-1} \rangle \hat{c}_{j} + pe^{-\lambda} \langle \hat{c}_{j-1} \rangle \hat{c}^\dagger_{j}-q\hat{n}_{j}(1-\langle \hat{n}_{j-1}\rangle ) \quad&\text{ for } c > 1 \nonumber\\
\hat{L}(j) &= -\alpha(\mathbbm{1}-\hat{n}_1) + \alpha e^{-\lambda} \hat{c}_1^\dagger + \gamma e^{\lambda}\hat{c}_1 -\gamma \hat{n}_1  \quad&\text{ for } c = 1 \nonumber\\
\hat{R}(j) &= -q\langle \hat{n}_{j+1}\rangle(\mathbbm{1}-\hat{n}_{j}) + pe^{-\lambda} \langle \hat{c}^\dagger_{j+1} \rangle \hat{c}_{j} + qe^{\lambda} \langle \hat{c}_{j+1} \rangle \hat{c}^\dagger_{j}-p\hat{n}_{j}(1-\langle \hat{n}_{j+1}\rangle ) \quad&\text{ for } c < L/c_L%
\nonumber\\
\hat{R}(j) &= -\nu(\mathbbm{1}-\hat{n}_L) + \nu e^{\lambda} \hat{c}_L^\dagger + \beta e^{-\lambda}\hat{c}_L -\beta \hat{n}_L \quad&\text{ for } c = L/c_L\nonumber
\end{align}
\end{widetext}
This treatment ensures that all cluster based eigenvalue problems can be solved separately
using only expectation values to estimate the couplings between clusters. The latter couplings are calculated separately at the end of each self-consistent step. This is continued until the averages do not change. Once the MF calculations converge we will obtain the required GDF $\langle \tilde{\Xi}|$ needed to construct the generator of auxiliary dynamics, similar to the case for the single site MF. However, now the $\xi_c(n_{(c-1)c_L+1},...,n_{cc_L})$ involve multiple sites and thus the proposal matrix must be updated accordingly to distinguish between inter- and intra- cluster states. For instance, for a cluster of size $c_L = 2$ the intra-cluster hopping matrix for hops within a cluster is given by,
\begin{align}
\tilde{u}_c &= \left[ 
\begin{array}{cccc}
1 & 0 & 0 & 0\\
 0 & 1-qdt & p e^{-\lambda} \frac{\xi_c(0,1)}{\xi_c(1,0)}& 0\\
 0 & qe^{\lambda} \frac{\xi_{c}(1,0)}{\xi_{c}(0,1)} & 1-pdt & 0\\ 
 0 & 0 & 0 & 1 
\end{array} 
\right]
\label{scu:eq1}
\end{align}
For inter-cluster hopping the two matrices $\tilde{u}_{c}$ and $\tilde{u}_{c+1}$ must be combined and the hopping between edge sites $n_{cc_L}$ and $n_{(c)c_L+1}$ will depend on the state configuration of the two clusters. For e.g. the transition matrix elements given by $P(n_{cc_L-1}, n_{cc_L}; n_{cc_L+1},n_{cc_L+2} \rightarrow 
n_{cc_L-1}, n_{cc_L}; n_{cc_L+1},n_{cc_L+2})$ are:
\begin{align}
P(i, 0; 0,j \rightarrow i, 0; 0,j) &= 1 \nonumber\\
P(i, 1; 0,j \rightarrow i, 0; 1,j) &= pe^{-\lambda} \frac{\xi_c(i,0)\xi_{c+1}(1,j)}{\xi_c(i,1)\xi_{c+1}(0,j)} \nonumber\\
P(i, 1; 0,j \rightarrow i, 1; 0,j) &= 1 - pdt \nonumber\\
P(i, 0; 1,j \rightarrow i, 1; 0,j) &= qe^{\lambda} \frac{\xi_c(i,1)\xi_{c+1}(0,j)}{\xi_c(i,0)\xi_{c+1}(1,j)} \nonumber\\
P(i, 0; 1,j \rightarrow i, 0; 1,j) &= 1 - qdt \nonumber\\
P(i, 1; 1,j \rightarrow i, 1; 1,j) &= 1 
\end{align}  
This can be generalized to other cluster sizes.

\subsection{2D Cluster Approximation}
For the 2D WASEP problem that we consider in the main text, the cluster approach outlined above for 1D systems can be generalized to 2D. The tilted operator for this problem is given by,
\begin{widetext}
\vspace{-10pt}
\begin{align}
\Wun_{\lambda_x,\lambda_y} = \sum_{i=1}^{L_x}\sum_{j=1}^{L_y} & p_x e^{-\lambda_x} \hat{c}^\dagger_{i+1,j}\hat{c}_{i,j} - p_x\hat{n}_{i,j}(\mathbbm{1}-\hat{n}_{i+1,j}) + q_xe^{\lambda_x}\hat{c}^\dagger_{i-1,j}\hat{c}_{i,j} - q_x\hat{n}_{i,j}(\mathbbm{1}-\hat{n}_{i-1,j}) \nonumber\\
& p_y e^{-\lambda_y} \hat{c}^\dagger_{i,j+1}\hat{c}_{i,j} - p_y\hat{n}_{i,j}(\mathbbm{1}-\hat{n}_{i,j+1}) + q_ye^{\lambda_y}\hat{c}^\dagger_{i,j-1}\hat{c}_{i,j} - q_y\hat{n}_{i,j}(\mathbbm{1}-\hat{n}_{i,j-1}),
\label{eq:WASEP}
\end{align}
\end{widetext}
where $p_x, q_x, p_y, q_y$ are transition rates associated with hopping along $\pm x, \pm y$ directions. Additionally for this model we use periodic boundary conditions. Here, $(L_x,L_y)$ are the number of sites in the respective directions. Notice that the biasing parameter is a vector $\bm{\lambda} = (\lambda_x,\lambda_y)$. Additionally to accommodate the extra dimension, operators are now indexed via the coordinates $(i,j)$. Much like the 1D case the MF solutions can be constructed as before except now the coupling terms are defined for the perimeter of the cluster at $(c_x,c_y)$. Suppose there are $(c_L^x,c_L^y)$ sites per cluster; then in explicit terms the cluster MF tilted operator $V_{c_x,c_y}$ is given by:
\begin{widetext}
\vspace{-20pt}
\begin{align}
V_{c_x,c_y} = \sum_{\substack{i=c_L^x(c_x-1)+1\\ j=c_L^y(c_y-1)+1}}^{\substack{c_L^xc_x-1, c_L^yc_y-1}}  &  p_x e^{-\lambda_x} \hat{c}^\dagger_{i+1,j}\hat{c}_{i,j} - p_x\hat{n}_{i,j}(\mathbbm{1}-\hat{n}_{i+1,j}) + q_xe^{\lambda_x}\hat{c}^\dagger_{i,j}\hat{c}_{i+1,j} - q_x\hat{n}_{i+1,j}(\mathbbm{1}-\hat{n}_{i,j}) \nonumber\\
 & + p_y e^{-\lambda_y} \hat{c}^\dagger_{i,j+1}\hat{c}_{i,j} - p_y\hat{n}_{i,j}(\mathbbm{1}-\hat{n}_{i,j+1}) + q_ye^{\lambda_y}\hat{c}^\dagger_{i,j}\hat{c}_{i,j+1} - q_y\hat{n}_{i,j+1}(\mathbbm{1}-\hat{n}_{i,j}) \nonumber\\
  + \hat{L}(i = (c_x-&1)c_L^x+1,c_y) + \hat{R}(i = c_xc_L^x,c_y) + \hat{T}(c_x, j = c_y c_L^y) + \hat{B}(c_x,j = (c_y-1)c_L^y+1) \nonumber\\ 
\end{align}
\vspace{-20pt}
\end{widetext}
\begin{widetext}
where
\begin{align}
\hat{L}(i,c_y) &= \sum_{j=c_L^y(c_y-1)+1}^{c_L^yc_y-1} 
p_x e^{-\lambda_x} \hat{c}^\dagger_{i,j}\langle\hat{c}_{i-1,j}\rangle - p_x\langle\hat{n}_{i-1,j}\rangle(\mathbbm{1}-\hat{n}_{i,j}) +
q_xe^{\lambda_x}\langle\hat{c}^\dagger_{i-1,j}\rangle\hat{c}_{i,j} - q_x\hat{n}_{i,j}(\mathbbm{1}-\langle\hat{n}_{i-1,j}\rangle) \nonumber
\end{align}
\begin{align}
\hat{R}(i,c_y) &= \sum_{j=c_L^y(c_y-1)+1}^{c_L^yc_y-1} 
p_x e^{-\lambda_x} \langle\hat{c}^\dagger_{i+1,j}\rangle\hat{c}_{i,j} - p_x \hat{n}_{i,j}(\mathbbm{1}-\langle\hat{n}_{i+1,j}\rangle) +
q_xe^{\lambda_x}\hat{c}^\dagger_{i,j}\langle\hat{c}_{i+1,j}\rangle - q_x\langle\hat{n}_{i+1,j}\rangle(\mathbbm{1}-\hat{n}_{i,j}) \nonumber\\
\hat{B}(c_x,j) &= \sum_{i=c_L^x(c_x-1)+1}^{c_L^xc_x-1} 
p_y e^{-\lambda_y} \hat{c}^\dagger_{i,j}\langle\hat{c}_{i,j-1}\rangle - p_y\langle\hat{n}_{i,j-1}\rangle(\mathbbm{1}-\hat{n}_{i,j}) +
q_ye^{\lambda_y}\langle\hat{c}^\dagger_{i,j-1}\rangle\hat{c}_{i,j} - q_y\hat{n}_{i,j}(\mathbbm{1}-\langle\hat{n}_{i,j-1}\rangle) \nonumber\\
\hat{T}(c_x,j) &= \sum_{i=c_L^x(c_x-1)+1}^{c_L^yc_x-1} 
p_y e^{-\lambda_y} \langle\hat{c}^\dagger_{i,j+1}\rangle\hat{c}_{i,j} - p_y \hat{n}_{i,j}(\mathbbm{1}-\langle\hat{n}_{i,j+1}\rangle) +
q_ye^{\lambda_y}\hat{c}^\dagger_{i,j}\langle\hat{c}_{i,j+1}\rangle - q_y\langle\hat{n}_{i,j+1}\rangle(\mathbbm{1}-\hat{n}_{i,j}) \nonumber
\end{align}
\end{widetext}
The operators $\hat{L}$, $\hat{R}$, $\hat{B}$ and $\hat{T}$ are the border terms for the cluster at $(c_x,c_y)$. They contain the mean-field coupling parameters that are enclosed with $\langle ... \rangle$. These are determined self-consistently much as in the 1D case. Finally the determined cluster states can then be used to construct the transformed tilted operator from which moves can be proposed.

\section{Calculation details for results in the main text}
\begin{figure}[b]
\begin{center}
\includegraphics[width=8.5cm]{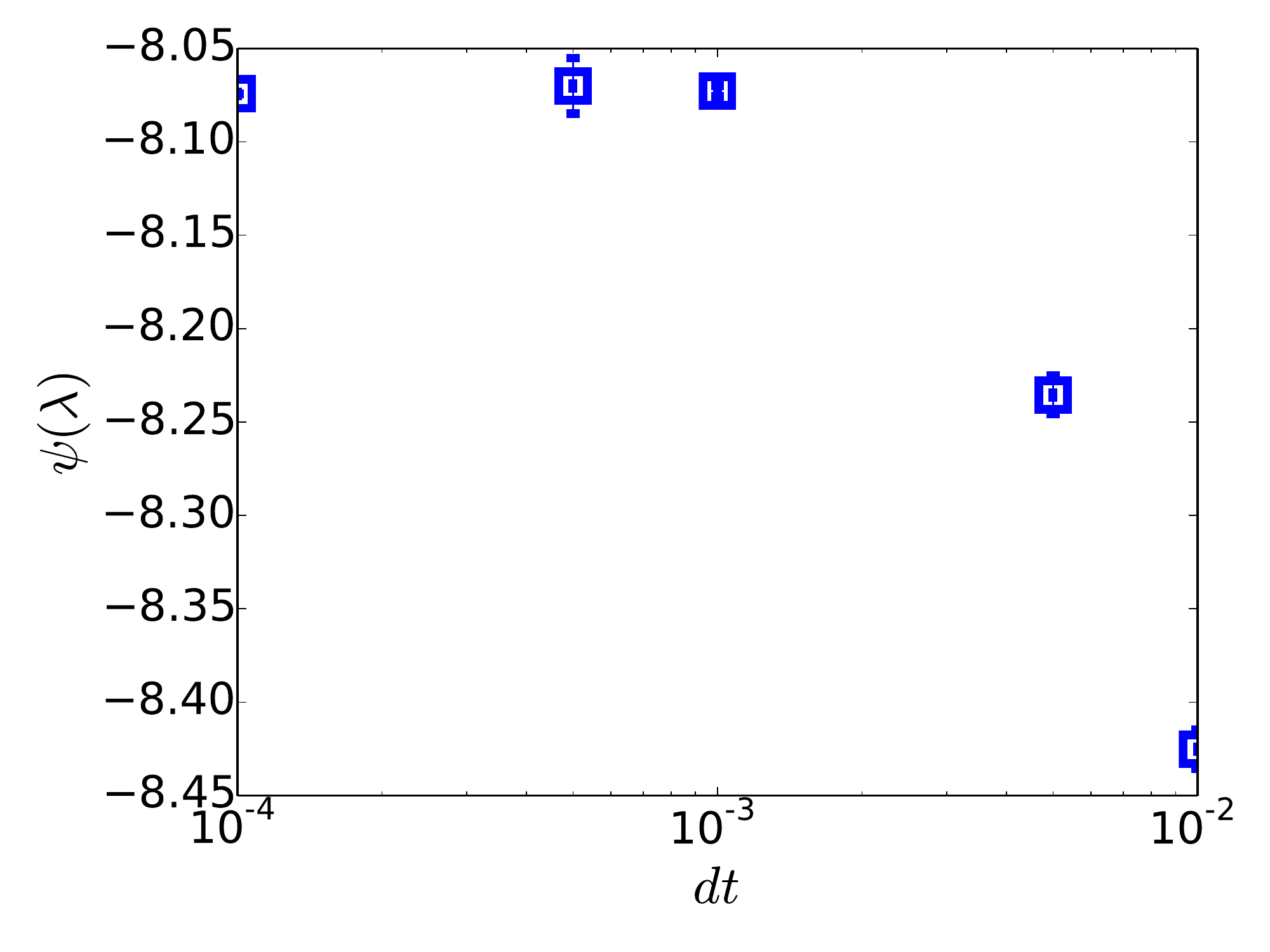}
\caption{Convergence of the 2D SEP DMC simulations with integration time step for $\lambda=-6$.}
\label{Fi:1}
\end{center}
\vspace{-20pt} 
\end{figure}

\begin{figure}[b]
\begin{center}
\includegraphics[width=8.5cm]{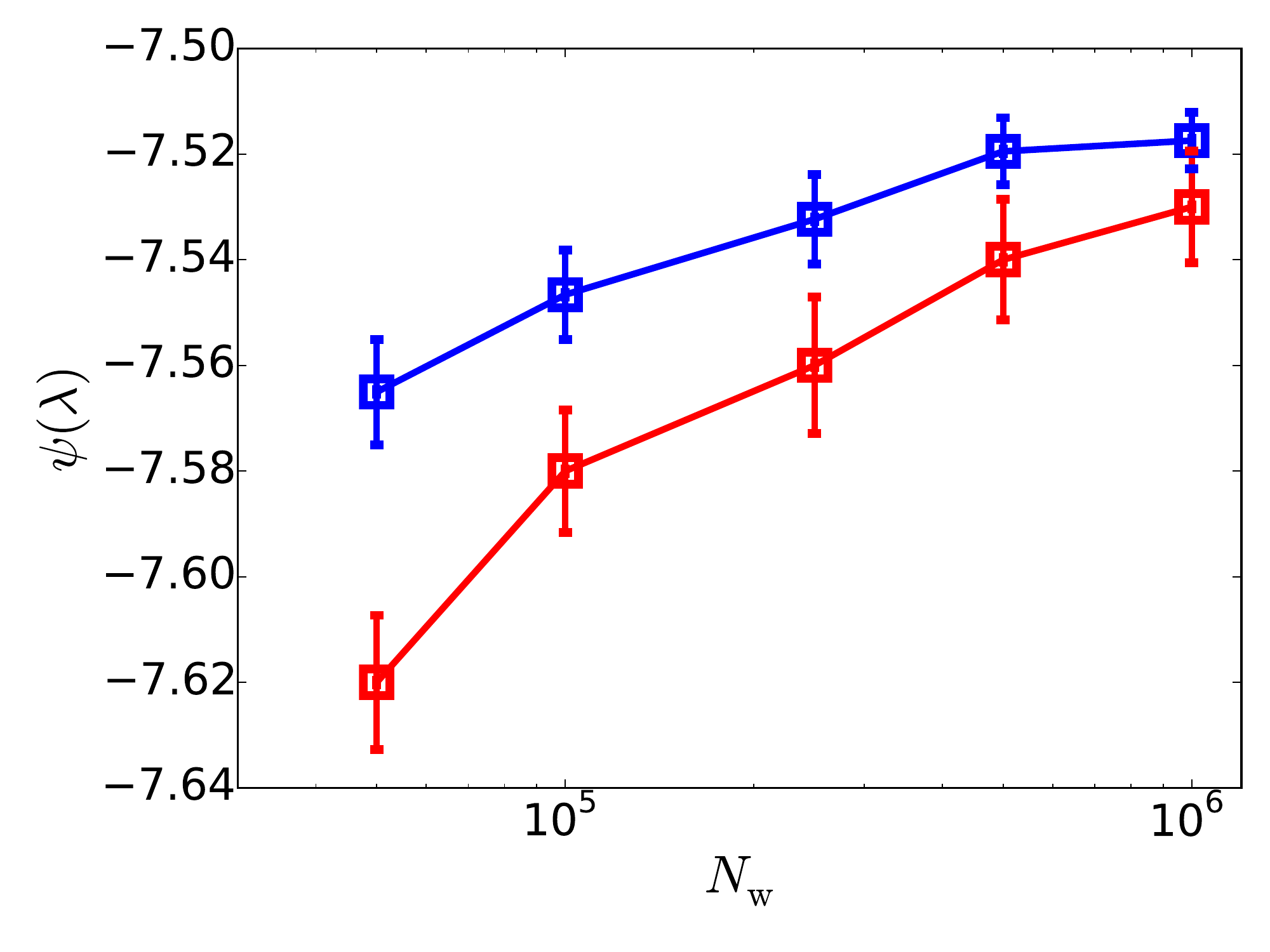}
\caption{Systematic error associated with walker population for 2D SEP DMC simulations for the large deviation function. Shown in red are the results for no importance sampling and in blue the results of the cluster GDF. Both were computed at $\lambda = -5.0$.}
\label{Fi:2}
\end{center}
\vspace{-20pt}
\end{figure}

\begin{figure}[b]
\begin{center}
\includegraphics[width=8.cm]{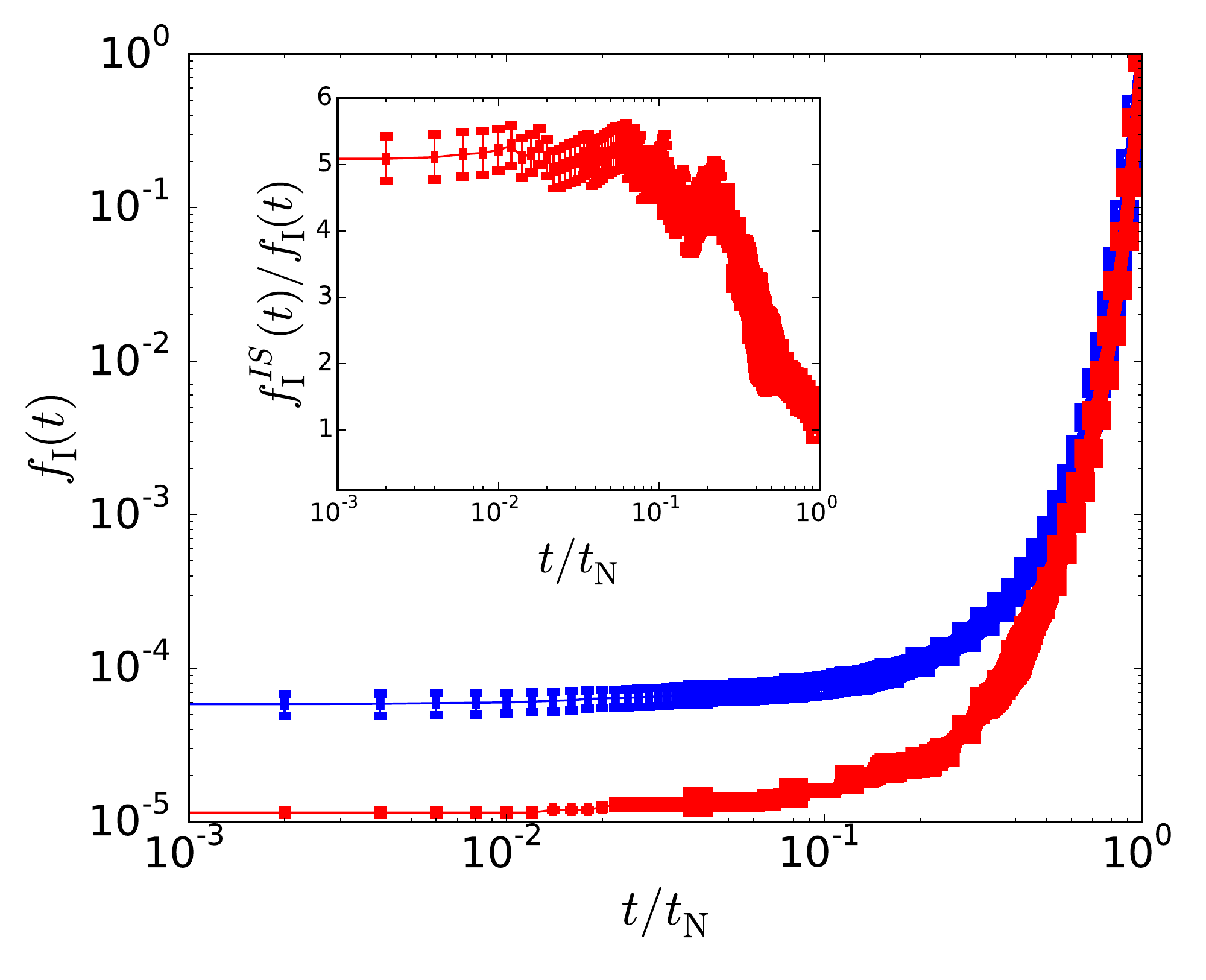}
\caption{Fraction of independent walkers $f_I(t)$ as a function of simulation time for $\lambda = -5.0$ using GDF (blue) and not using GDF (red). Inset shows the ratio of fraction of independent walkers with GDF and without as a function of time.}
\label{Fi:3}
\end{center}
\vspace{-20pt} 
\end{figure}

\emph{Driven brownian motion}: All of the calculations on the driven brownian walker were accomplished with a second order stochastic Runge-Kutta integrator with a timestep of 0.01. Observation times of 20.0 were needed to converge the calculations and branching steps attempted every 0.05 unit of time.\\ 

\emph{1D SEP}: For the 1D system consisting of $L = 8$ sites, clusters were made using $c_L = 1,2,4$ sites. All calculations were done using $\Nw = 2\times 10^4$ walkers, with a time step $dt = 0.001$, observation time $\tobs = 1$ with branching occurring at intervals of time $t_\mathrm{int} = 0.01$. In the main text, we have used calculations done with cluster size of $0$ (``zeroth-order'' MF) to indicate a sampling strategy where the exponential $e^{-\lambda\mathcal{O}}$ has been absorbed directly into the proposed Monte Carlo moves. This essentially means $\xi_i = 1$ for equations \ref{su:eq1}-\ref{su:eq3}. We note that this last way of sampling should always be used if no higher order MF solutions are available. For the 2D WASEP system we discuss next, the bare sampling strategy is to at least use a zeroth-order MF. Generating trajectories from the unbiased distribution, i.e., without directly incorporating the exponential in the proposed Monte Carlo moves makes it impossible to converge calculations for the range of $\lambda$ we explore and the number of walkers we deploy.\\ 

\emph{2D WASEP}: The 2D WASEP system discussed in the main text is a generalization of the 1D model to an $L\times L$ lattice with  periodic boundary conditions. Along the principal direction, $x$, particles are subject to  biased hopping rates that are scaled by the length of the system $p_x = e^{-E/L_x}/2$ along $x$ and $q_x = e^{E/L_x}/2$ along $-x$ with $L_x = 12$ and $E = 10$. Along the $y$ direction particles diffuse symmetrically, with rates $p_y = q_y = 0.5$ \cite{tizon2016order}. These calculations are resource intensive so a careful study of convergence and statistical properties is needed to be able to compute the susceptibility $\chi(\lambda)$. 

Towards this end we determine the time step error associated with Trotterization as shown in Fig. \ref{Fi:1} for the largest $\lambda$ value that we wish to access, since this sets the upper-bound on the error. We find that $dt = 0.001$ is sufficient to converge the error.  

The second major source of error in these calculations is the systematic error due to finite walker population. Since the variance grows exponentially with $|\lambda|$ \cite{ray2017importance}, it is sufficient to determine the largest number of walkers needed for maximal $|\lambda|$ we use in our simulations. Fig. \ref{Fi:2} shows the convergence of $\psi(\lambda)$ with $\Nw$ for $\lambda = -5.0$ both using and not using importance sampling. It is evident that using auxiliary dynamics, even with $\Nw = 5\times 10^5$ our results would have been sufficiently converged. Comparing against calculations without auxiliary dynamics we have $\sim 4$-fold increase in efficiency mirroring the ratio of independent walkers of Fig. 3c in the main text. Despite the mean field GDF employed for this model not being particularly good at large $|\lambda|$, as the traveling wave state is not well described by a small single cluster, we still get a factor of 2-4 reduction in the required number of walkers to converge results (c.f. Fig. 3c in main text).
 
Following this analysis, we used $\Nw = 5\times 10^5$ walkers for $\lambda \le -3.0$ with $\tobs = 100$ where branching was done every $t_\mathrm{int} = 0.01$. 
For $\lambda > -3.0$ simulations used $\Nw = 10^6$ walkers with $\tobs = 72$ and $t_\mathrm{int} = 1.4$. In Fig. 3c of the main text the ratio of independent clones was determined using independent clone counts at time $t = 0.32\tobs$. 
Fig. \ref{Fi:3} shows a comparison of the fraction of independent walkers, $f_\mathrm{I}(t)$, along the entire $\tobs$ trajectory for $\lambda = -5.0$. It highlights the importance of using GDF for sampling purposes to ensure that that there are enough uncorrelated contributions to the estimator.   

\section{Connection with the iterative feedback method of Nemoto et al}

In their work, Nemoto et al. have outlined an iterative feedback control algorithm by which they attempt to construct a potential to affect an importance sampling of the large deviation function\cite{Takahiro2016,nemoto2017finite}. These papers illustrate their method via two models: diffusion of a particle in a quartic potential and the FA model\cite{fredrickson1984kinetic}. The basic idea in the context of a lattice model is to modify the transition rate $w(\mathcal{C}\rightarrow\mathcal{F}[\mathcal{C}])$ between configurations $\mathcal{C}$ and $\mathcal{F}[\mathcal{C}] $ via a auxilary potential $U(\mathcal{C})$ as
\begin{align}
\tilde{w}(\mathcal{C}\rightarrow\mathcal{F}[\mathcal{C}]) = e^{-\lambda}w(\mathcal{C}\rightarrow\mathcal{F}[\mathcal{C}])e^{\frac{1}{2}[U(\mathcal{C})-U(\mathcal{F}[\mathcal{C}])]}.
\end{align}
where they biased on the activity, or number of configurational changes. 
Here $\mathcal{F}[\mathcal{C}]$ is a function that maps the current configuration $\mathcal{C}$ to a new configuration and $\lambda$ is a counting field conjugate to the activity. Typically Nemoto et al. parametrize $U(\mathcal{C})$ iteratively using the criteria that in the optimum Doob transformed dynamics the distribution of order parameter values computed over the full simulation time -- the ``average distribution" and those constructed from the final ``end distribution" must be identical. Of the models they consider the more complicated many-particle-interacting simulations for the FA case is done by setting $U(\mathcal{C})-U(\mathcal{F}_i[\mathcal{C}]) \equiv u_d(n_{i-d},...,n_i,...,n_{i+d})$. Here $\mathcal{F}_i$ simply constitutes a spin-flip on site $i$ and $\mathcal{C}$ represents a 1D lattice of spins. 

Nemoto et al. alluded to the relationship between their 
%
 potential and the dominant left eigenvector of the tilted operator, but did not give an explicit expression. From the formalism presented in the text it is evident that 
\begin{align}
e^{\frac{1}{2}[U(\mathcal{C})-U(\mathcal{F}[\mathcal{C}])]} = \frac{\tilde{\Xi}(\mathcal{F}[\mathcal{C}])}{\tilde{\Xi}(\mathcal{C})},
\end{align}
where $\tilde{\Xi}(\mathcal{C})$ is the GDF obtained by approximating the left eigenvector of the tilted operator, as we have  discussed in the main text. This connection  simplifies 
the importance sampling in the sense that GDFs with larger overlap with the exact left eigenvector of the tilted operator immediately
lead to better sampling. The iterative mechanism outlined by Nemoto et al. may be seen as an analog of the variation of parameters
needed to determine a parametrized GDF. However, since the variational determination of the GDFs is carried out outside of the dynamics itself, we expect it
to be numerically less intensive in most situations than the iterative feedback algorithm.


\bibliography{v1}